\documentclass{edp-conf}
\usepackage{graphicx}

\begin{document}

\TitreGlobal{SF2A 2005}

\title{Length of day and polar motion, with respect to temporal variations of the Earth gravity field}

\author{Bourda, G.$^{1,}$}\address{Observatoire de Paris, SYRTE/UMR8630, France, email: Geraldine.Bourda@obspm.fr}\address{Vienna University of Technology, Institute of Geodesy and Geophysics, Austria}

\runningtitle{LOD and Gravity Field variations}

\setcounter{page}{237}
\index{Bourda, G.}

\maketitle

\begin{abstract}
The masses distribution inside the Earth governs the behaviour of the rotation axis in the Earth (polar motion), as well as the Earth rotation rate (or equivalently, length of day). This masses distribution can be measured from space owing to artificial satellites, the orbitography of which provides the Earth gravity field determination. Then, the temporal variations of the Earth gravity field can be related to the variations of the Earth Orientation Parameters (EOP) (with the Inertia Tensor). Nowadays, owing to the satellite laser ranging (SLR) technique and to the new gravimetric satellite missions (as CHAMP or GRACE), the temporal variations of the low degree coefficients of the Earth gravity field (i.e. Stokes coefficients) can be determined. This paper is one of the first study using gravity variations data in the equations already established (e.g. Lambeck 1988) and linking the variations of the length of day and of the $C_{20}$ Stokes coefficient (or, linking the polar motion and the $C_{21}$ and $S_{21}$ coefficients). This paper combines the Earth rotation data (mainly obtained from VLBI and GPS measurements) and the Earth gravity field variations ones (e.g. Lageos I and II data, or GRACE data), in order to complete and constrain the models of the Earth rotation. The final goal is a better Earth global dynamics understanding, which possible application can be the constraint on the couplings into the Earth system (e.g. core-mantle couplings). 
\end{abstract}

\section{Introduction}
In this paper, we compare the mass terms of the Earth Rotation Parameters (ERP) series, length of day (LOD) and polar motion, obtained from various techniques. We use (i) on one hand the IERS (International Earth Rotation and refrence systems Service) C04 series (obtained combining mainly VLBI and GPS measurements), and (ii) on the other hand the gravity field variations data (CNES/GRGS gravity field degree~2-coefficients, obtained from Lageos~I and II positionning measurements: for further explanation see Bourda \& Capitaine 2004).

\section{Computations}
The exces in the length of day $\Delta(LOD)$ (with respect to 86400~s) can be related to the temporal variations of the Stokes cofficient $C_{20}$ (Bourda~2004a, Bourda~2004b):
\begin{equation}\label{eq:LOD}
\frac{\Delta(LOD)_{mass} (t)}{86400~s} = -\frac{2}{3 ~C_m}~M~{R_e}^2 ~\Delta C_{20} (t)  
\end{equation}
where $C_m$ is the third moment of inertia of the Earth's mantle, $M$ is the mass of the Earth, $R_e$ is the mean equatorial radius of the Earth, and $\Delta C_{20}$ is related to the motion of masses from the equator till the northern or the southern hemisphere (and the opposite).

The excitation $\chi$ of the polar motion $p = x_p - i~y_p$, where $x_p$ and $y_p$ are the components of the rotation axis in the Earth, can be theoretically related to the degree 2 and order 1 coefficients of the Earth gravity field (Bourda~2004a, Bourda~2004b):
\begin{equation}\label{eq:PM}
\chi_{mass} (t) = - \frac{k_0}{k_0-k_2} ~\frac{M~{R_e}^2}{C_m-A_m} ~\left( C_{21} (t) + i~S_{21} (t) \right) 
\end{equation}
where $\frac{k_0}{k_0-k_2}=1.43$ (Barnes et al.~1983), and $\chi_G = p + \frac{i}{\sigma_0} ~\dot{p}$, with $\chi_G$ the geodetic polar motion excitation (i.e.~the observed one) and $\sigma_0$ the Chandler pulsation.

\section{Discussion}
Zonal tides, motion terms, oceanic tides and atmospheric angular momentum effects were removed from the IERS C04 series for valuable comparisons. We noticed a good agreement between LOD and polar motion exciation obtained from (i) gravity field variations data, and (ii) classical geodetic measurements. But it still remains some critical points. For example, the correlation between the IERS C04 LOD series and the one from $\Delta C_{20}$ residual data (i.e. the measurements, without solid Earth tides, atmospheric pressure effects, and oceanic tides) is poor: 0.17. But this can come from the using of various models for the same effect (e.g. the atmospheric pressure part) in these two methods. Further investigations will focuss on the errors made removing atmospheric winds from LOD data (which is the main effect), and on the hypothesis $\Delta Tr(I)=0$ (Bourda 2004a) assumed computing Eq.~(\ref{eq:LOD}) (which is perhaps not valuable using the $\Delta C_{20}$ residuals).


\end{document}